\documentclass[twocolumn, amsmath, amssymb, aps]{revtex4}

\usepackage{graphicx}
\usepackage{float}
\usepackage{dcolumn}
\usepackage{subfigure}
\usepackage{amsthm}
\usepackage{tensor}
\usepackage{color}
\usepackage[all]{xy}
\usepackage{tikz}
\usepackage{dsfont}
\usepackage{times,txfonts}
\usetikzlibrary{positioning}
\usepackage{braket}

\theoremstyle{definition}

\newcommand{\inn}[2]{\left\langle{#1}|{#2}\right\rangle}

\begin{document}

\title{A variational quantum algorithm for tackling multi-dimensional Poisson equations with inhomogeneous boundary conditions}

\author{Minjin Choi}
\affiliation{
Center for Quantum Information R\&D, Korea Institute of Science and Technology Information, Daejeon 34141, Republic of Korea
}

\author{Hoon Ryu} 
\email{elec1020@kumoh.ac.kr}
\affiliation{
Department of Artificial Intelligence Engineering, Kumoh National Institute of Technology, Gumi, Gyeongsangbuk-do 39177, Republic of Korea
}

\date{\today}

\begin{abstract}
We design a variational quantum algorithm to solve multi-dimensional Poisson equations with mixed boundary conditions that are typically required in various fields of computational science. Employing an objective function that is formulated with the concept of the minimal potential energy, we not only present in-depth discussion on the cost-efficient \& noise-robust design of quantum circuits that are essential for evaluation of the objective function, but, more remarkably, employ the proposed algorithm to calculate bias-dependent spatial distributions of electric fields in semiconductor systems that are described with a two-dimensional domain and up to 10-qubit circuits. Extending the application scope to multi-dimensional problems with mixed boundary conditions for the first time, fairly solid computational results of this work clearly demonstrate the potential of variational quantum algorithms to tackle Poisson equations derived from physically meaningful problems.
\end{abstract}
\maketitle

\section{Introduction}
\label{sec: introduction}

The Poisson equation is a second-order partial differential equation that plays a significant role in analyzing physical phenomena in various fields of computational science such as fluid dynamics~\cite{Batchelor00, Kang2021}, nanoelectronics~\cite{Ryu2015, Ryu2016, Ryu2022, Kang2021_2}, computational electromagnetics~\cite{Griffiths23, Shan2020}. In general, it is not easy to determine the solution of a Poisson equation analytically, so computer-based calculations are commonly employed with aids of numerical analysis, by transforming the master problem into a linear system problem $A \mathbf{v} = \mathbf{b}$ where $\mathbf{v}$ represents the solution of a Poisson system matrix $A$ given by a right-hand-side (RHS) $\mathbf{b}$~\cite{Morton05}. Consequently, the computational expense of a Poisson equation problem directly depends on the degrees of freedom (DOFs) of the system matrix that would generally become larger if we need to increase the accuracy of numerical solutions or to handle larger domains in a real space. Recently, quantum computing has obtained huge attention due to the potential of solving some computational problems more efficiently than its classical counterpart by leveraging quantum phenomena \cite{Lundow2022, Kim2023, Harrow17, Yuri2024, Harrow09, Shor1994}, and, in particular, Harrow $et$ $al.$ proposed the Harrow-Hassidim-Lloyd~(HHL) algorithm that can achieve exponential speed-up relative to classical algorithms for solving linear system problems~\cite{Harrow09}. Yudong $et$ $al.$ demonstrated that approximate solutions of Poisson equations can be obtained based on the HHL algorithm~\cite{Cao13}, but its effectiveness in terms of computing cost hinges on fault-tolerant quantum computing that is not tangible yet. 

It is fair to say that the current status of quantum computing is in the noisy intermediate-scale quantum~(NISQ) era~\cite{Preskill18}, where computing devices are subject to substantial noises and thus are limited in the number of quantum bits (qubits). Variational quantum algorithms~(VQAs)~\cite{Cerezo21}, which are executable in NISQ devices through a hybrid utilization of quantum and classical computing, have led to significant efforts to explore their utility in diverse numerical applications such as determination of low-lying eigenpairs of symmetric matrices \cite{Perozzo2014} and solutions of linear systems \cite{Bravo23}, combinatorial optimization \cite{Morales2020}, and so on. VQAs dedicated to the Poisson equation have been also reported, showing their capability for securing solutions of simple one-dimensional problems~\cite{Liu21, Sato21}. Rigorous studies on their utility for the Poisson equation, however, are still necessary since physically meaningful problems are typically represented with multi-dimensional domains and mixed boundary conditions \cite{Ryu2015, Kang2021, Kang2021_2}, motivating comprehensive investigations with more complicated scenarios to examine the effectiveness of VQAs for prediction of physical phenomena with Poisson equations.

So in this work, we design a VQA that can be useful to calculate multi-dimensional problems of inhomogeneous boundaries with a Poisson equation. To achieve cost-efficiency and noise-robustness of the algorithm, we begin the design process with adoption of strong points of the two previous studies that not only focus on one-dimensional problems, but are limited by impracticality in the objective (cost) function~\cite{Liu21} and the circuit decomposition~\cite{Sato21}. In more detail, we employ the cost function derived upon the concept of the minimal potential energy by Sato {\em et al.}~\cite{Sato21} that is much simpler than the one employed by Liu {\em et al.}~\cite{Liu21}. For evaluation of the cost function, however, Sato {\em et al.} utilize quantum circuits involving multi-controlled $X$ logics that are not quite affordable in the NISQ regime, so here we design evaluation circuits with only one- \& two-qubit gates as Liu {\em et al.} have done. The algorithm designed in this work can evaluate the cost function with circuits of polynomial complexities, and, more remarkably, the number of circuits required to get the cost function is independent of the dimensionality of target problems.

Finally, the practicality of our algorithm is carefully examined with two-dimensional problems including the calculation of potential energy distributions in a silicon-based structure that are manipulated by electrical biases imposed on multiple electrodes. All the numerical simulations are conducted with the PENNYLANE software development kit~\cite{Pennylane} to emulate a gate-based quantum computer, and the optimizer based on L-BFGS-B algorithm~\cite{Liu89} is used to classically update ansatz parameters. Overall results of this work clearly indicate that multi-dimensional Poisson equations of mixed boundary conditions can be solved with fairly sound accuracy, revealing the potential of VQA to address physically meaningful problems.
\section{Methods}
\label{Sec: methods}

\subsection{The cost function employed to solve Poisson equations}
\label{Cost function}
Let us consider the $d$-dimensional Poisson equation
\begin{equation}
-\Delta u(\mathbf{x}) = f(\mathbf{x}), 
\quad \mathbf{x} \in (0, 1)^{d},
\label{eq:01}
\end{equation}
where $\Delta$ is the Laplace operator and $f$ is a given smooth function~\cite{Evans22}. Equation (\ref{eq:01}) can be then numerically transformed into a linear system problem
\begin{equation}
A \mathbf{v} = \mathbf{b},
\label{eq:02}
\end{equation}
where $A$ is a $m^{d} \times m^{d}$ system matrix describing the computational domain, $\mathbf{b}$ is a $m^{d} \times 1$ vector that represents a perturbation given by the function $f$, and $m$ is a number of grids employed for discretization of the domain along each axis. The solution $\mathbf{v}$ of a linear system problem in Equation (\ref{eq:02}) then becomes the approximate solution of the Poisson equation. When a $d$-dimensional computational domain is discretized with a finite difference method~\cite{Morton05}, the Poisson system matrix $A$ becomes
\begin{eqnarray}
\label{mat_dPoisson}
A &=& A_{1} \otimes I^{(m)} \otimes \cdots \otimes  I^{(m)} + I^{(m)} \otimes A_{2} \otimes I^{(m)} \otimes \cdots  \otimes I^{(m)} \nonumber \\
&& + \cdots +   I^{(m)} \otimes \cdots \otimes I^{(m)} \otimes A_{d},
\label{eq:03}
\end{eqnarray}
where $I^{(m)}$ is a $m \times m$ identity matrix and $A_{i}$ is a $m \times m$ matrix representing the one-dimensional Poisson equation of the $i$-th axis. The one-dimensional Poisson matrix $A_{i}$ depends on the boundary condition, so, for example, it is given by
\begin{equation}
A_{D} \equiv 
\begin{pmatrix}
2 & -1 & 0 & \cdots & 0 & 0 & 0\\
-1 & 2 & -1 & \cdots & 0 & 0 & 0\\
0 & -1 & 2 & \cdots & 0 & 0 & 0\\
  & \vdots & & \ddots & & \vdots & \\
0 & 0 & 0 & \cdots & 2 & -1 & 0 \\
0 & 0 & 0 & \cdots & -1 & 2 & -1 \\
0 & 0 & 0 & \cdots & 0 & -1 & 2 
\end{pmatrix}
\in \mathbb{R}^{m \times m}
\label{eq:04}
\end{equation}
with Dirichlet (closed) boundaries, and
\begin{equation}
A_{N} \equiv 
\begin{pmatrix}
1 & -1 & 0 & \cdots & 0 & 0 & 0\\
-1 & 2 & -1 & \cdots & 0 & 0 & 0\\
0 & -1 & 2 & \cdots & 0 & 0 & 0\\
  & \vdots & & \ddots & & \vdots & \\
0 & 0 & 0 & \cdots & 2 & -1 & 0 \\
0 & 0 & 0 & \cdots & -1 & 2 & -1 \\
0 & 0 & 0 & \cdots & 0 & -1 & 1 
\end{pmatrix}
\in \mathbb{R}^{m \times m}
\label{eq:05}
\end{equation}
with Neumann (zero-derivative) boundaries. Then, a Poisson system matrix $A_{ND}$ that describes a two-dimensional domain having Neumann and Dirichlet boundaries along the $x_{1}$- and the $x_{2}$-axis, respectively, can be expressed as
\begin{equation}
\label{target_A}
A_{ND} = A_{N} \otimes I^{(m)} + I^{(m)} \otimes A_{D}.
\end{equation}

Since the Poisson matrix is always positive-definite, the linear system problem associated with Poisson equations can become an optimization problem. In more detail, for a positive-definite matrix $A$, the condition $A \mathbf{v}_{0}=\mathbf{b}$ is satisfied if and only if $g_{A, \mathbf{b}}(\mathbf{v})$ is minimized at $\mathbf{v}$ = $\mathbf{v}_{0}$, where $g_{A, \mathbf{b}}(\mathbf{v})$ is defined as
\begin{equation}
g_{A, \mathbf{b}}(\mathbf{v}) \equiv \frac{1}{2} \mathbf{v}^{\dagger} A \mathbf{v} - \frac{1}{2} \mathbf{b}^{\dagger} \mathbf{v} - \frac{1}{2} \mathbf{v}^{\dagger} \mathbf{b},
\label{eq:07}
\end{equation}
and, based on Equation (\ref{eq:07}), Sato \textit{et al.}~\cite{Sato21} introduced the cost function for VQA as
\begin{equation}
E(\pmb{\theta}) \equiv -\frac{1}{2}\frac{\left(Re\inn{\mathbf{b}}{\psi(\pmb{\theta})}\right)^{2}}{\bra{\psi(\pmb{\theta})}A\ket{\psi(\pmb{\theta})}},
\label{eq:08}
\end{equation}
where $\ket{\mathbf{b}}=\mathbf{b}/||\mathbf{b}||$, $\ket{\psi(\pmb{\theta})}$ is a parameterized ansatz state, and $Re\inn{\mathbf{b}}{\psi(\pmb{\theta})}$ is the real part of $\inn{\mathbf{b}}{\psi(\pmb{\theta})}$. We use VQA to find the set of optimal parameters $\pmb{\theta}_{opt}$ that minimizes the cost function $E(\pmb{\theta})$, and get a normalized solution $\ket{\psi(\pmb{\theta}_{opt})}$ if the optimization process is successful. Once $\ket{\psi(\pmb{\theta}_{opt})}$ is known, its norm can be approximated with $|r(\pmb{\theta}_{opt})|||\mathbf{b}||$ where $r(\pmb{\theta}_{opt})$ is given as
\begin{equation}
r(\pmb{\theta}_{opt})=\frac{Re\left<\mathbf{b}\big|\psi(\pmb{\theta}_{opt})\right>}{\bra{\psi(\pmb{\theta}_{opt})}A\ket{\psi(\pmb{\theta}_{opt})}}.
\end{equation}

If parameterized quantum gates in the ansatz circuit consist of only single-qubit rotations ($RX$, $RY$, $RZ$), the partial derivative of the cost function with respect to the $i$-th parameter of $\pmb{\theta}$, which is required by optimizations based on the L-BFGS-B algorithm~\cite{Liu89}, can be calculated as
\begin{align}
&\frac{\partial E(\pmb{\theta})}{\partial \theta_{i}} =
-\frac{1}{2}\frac{Re\inn{\mathbf{b}}{\psi(\pmb{\theta})} Re\inn{\mathbf{b}}{\psi(\pmb{\theta}+\pi_{i})}}{\bra{\psi(\pmb{\theta})}A\ket{\psi(\pmb{\theta})}} \nonumber \\
&+\frac{1}{2}\frac{\left(Re\inn{\mathbf{b} }{\psi(\pmb{\theta})}\right)^{2}  \bra{\psi(\pmb{\theta}), \psi(\pmb{\theta}+\pi_{i})}X \otimes A \ket{\psi(\pmb{\theta}), \psi(\pmb{\theta}+\pi_{i})}}
{\bra{\psi(\pmb{\theta})}A\ket{\psi(\pmb{\theta})}^{2}},
\label{eq:10}
\end{align}
where $\pmb{\theta}+\pi_{i}$ indicates adding $\pi$ to the $i$-th parameter of $\pmb{\theta}$, and $\ket{\psi(\pmb{\theta}),\psi(\pmb{\theta}+\pi_{i})}=(\ket{0}\otimes\ket{\psi(\pmb{\theta})}+\ket{1}\otimes\ket{\psi(\pmb{\theta}+\pi_{i})})/\sqrt{2}$ ($\ket{0}$ and $\ket{1}$ are single-qubit states).

\subsection{Estimation of the cost function and its partial derivatives with quantum circuits}
\label{Quantum circuits}

\begin{figure*}[t]
\includegraphics[width=\textwidth]{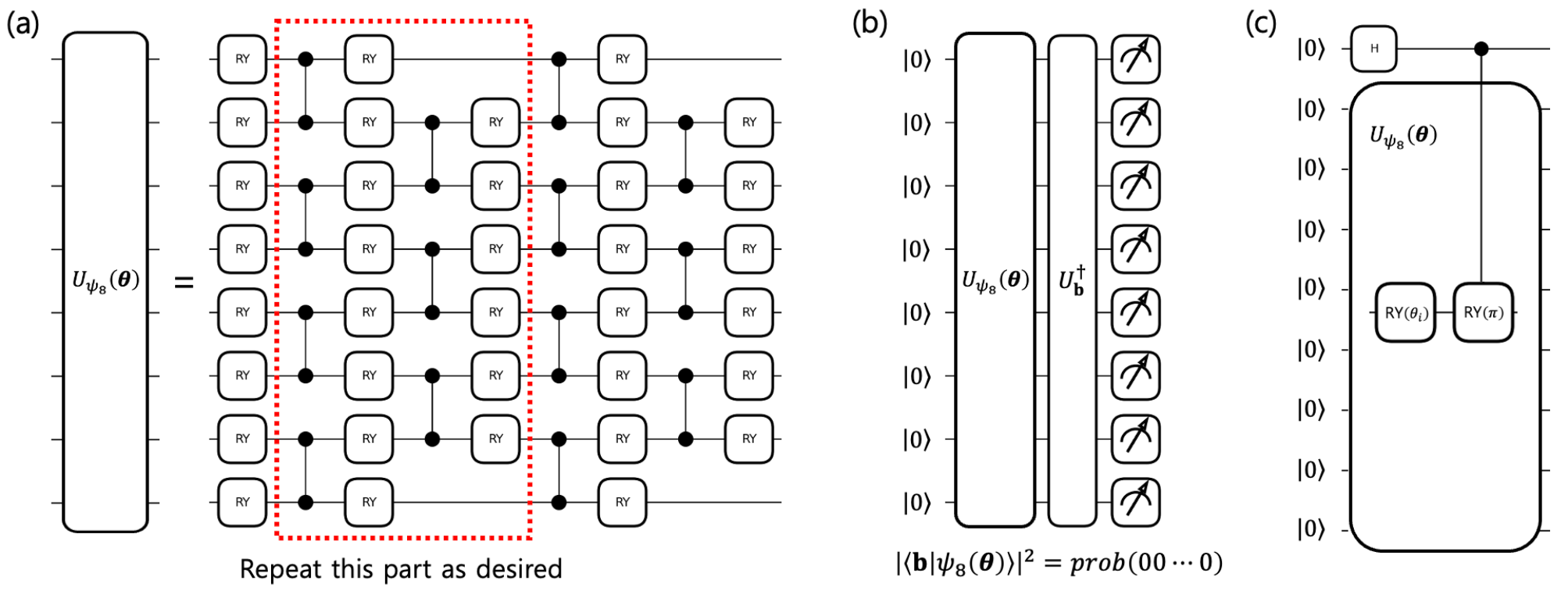}
\caption{\label{fig_ansatz}
\textbf{Quantum circuits for the case of (\textit{d} = 2, \textit{n} = 4) (I):} (a) The circuit employed to generate an ansatz state $\ket{\psi_{dn}(\pmb{\theta})}$. The unitary operator $U_{\psi_{dn}}(\pmb{\theta})$, which satisfies $U_{\psi_{dn}}(\pmb{\theta})\ket{0}^{\otimes dn}=\ket{\psi_{dn}(\pmb{\theta})}$, is composed of $RY$ gates and $CZ$ gates, ensuring that the ansatz state $\ket{\psi_{dn}(\pmb{\theta})}$ is a real-valued vector. Applying the quantum circuits in the red dotted line $p$ times introduces $dn+2p(dn-1)$ parameters. (b) The circuit used to evaluate the numerator $(Re\inn{\mathbf{b}}{\psi_{dn}(\pmb{\theta})})^{2}$ of the cost function $E(\pmb{\theta})$. With the unitary operator $U_{\mathbf{b}}$, the condition of $U_{\mathbf{b}}\ket{0}^{\otimes dn}=\ket{\mathbf{b}}$ is satisfied. (c) The circuit representing $\ket{\psi_{dn}(\pmb{\theta}), \psi_{dn}(\pmb{\theta}+\pi_{i})}$, which plays a key role in calculating the partial derivatives of $E(\pmb{\theta})$. The circuit is designed using the relation $RY(\pi)RY(\theta_{i})=RY(\theta_{i}+\pi)$ and an additional Hadamard gate ($H$).
}
\end{figure*}

Let's say each one-dimensional Poisson matrix in Equation (\ref{eq:03}) is described with $n$ qubits, $i.e.$ $m=2^{n}$, and we are handling a $d$-dimensional Poisson equation. The first step for estimation of the cost function $E(\pmb{\theta})$ is to secure the ansatz circuit. As the solution of a Poisson equation consists of real values, here we employ a hardware-efficient ansatz circuit as shown in Figure~\ref{fig_ansatz}(a), which always produces a real-valued state vector $\ket{\psi_{dn}(\pmb{\theta})}$, being composed of $RY$ and $CZ$ gates. In Figure~\ref{fig_ansatz}(b), we show a quantum circuit that is used to evaluate the numerator of $E(\pmb{\theta})$. Given that $\ket{\psi_{dn}(\pmb{\theta})}$ and $\ket{\mathbf{b}}$ are real-valued vectors, the numerator can be calculated as
\begin{equation}
\left(Re\inn{\mathbf{b}}{\psi_{dn}(\pmb{\theta})}\right)^{2}= 
\bra{\mathbf{0}} U_{\mathbf{b}}^{\dagger}U_{\psi_{dn}}(\pmb{\theta}) \ket{\mathbf{0}}^{2},
\end{equation}
where $\ket{\mathbf{0}}=\ket{0}^{\otimes dn}$, and unitary operators $U_{\psi_{dn}}(\pmb{\theta})$ and $U_{\mathbf{b}}$ satisfy conditions of $U_{\psi_{dn}}(\pmb{\theta})\ket{\mathbf{0}}=\ket{\psi_{dn}(\pmb{\theta})}$ and $U_{\mathbf{b}}\ket{\mathbf{0}}=\ket{\mathbf{b}}$, respectively. In addition, the numerator of the partial derivative of $E(\pmb{\theta})$ with respect to $\theta_{i}$ (Equation (\ref{eq:10})) can be also easily computed, because $Re\inn{\mathbf{b}}{\psi_{dn}(\pmb{\theta})} Re\inn{\mathbf{b}}{\psi_{dn}(\pmb{\theta}+\pi_{i})}$ is equivalent to
\begin{eqnarray}
\label{part of derivative}
2\left((\bra{+} \otimes \bra{\mathbf{b}}) \ket{\psi_{dn}(\pmb{\theta}), \psi_{dn}(\pmb{\theta}+\pi_{i})}\right)^{2} \nonumber \\
-\frac{1}{2}\inn{\mathbf{b}}{\psi_{dn}(\pmb{\theta})}^{2} - \frac{1}{2}\inn{\mathbf{b}}{\psi_{dn}(\pmb{\theta}+\pi_{i})}^{2},
\end{eqnarray}
where $\ket{+}=(\ket{0}+\ket{1})/\sqrt{2}$, and the state $\ket{\psi_{dn}(\pmb{\theta}),\psi_{dn}(\pmb{\theta}+\pi_{i})}$ can be evaluated using a circuit shown in Figure~\ref{fig_ansatz}(c) that is obtained by adding an ancilla qubit to the ansatz (Figure~\ref{fig_ansatz}(a)) with a controlled $RY(\pi)$ logic right after the $RY(\theta_{i})$ gate.

\begin{figure*}[t]
\includegraphics[width=\textwidth]{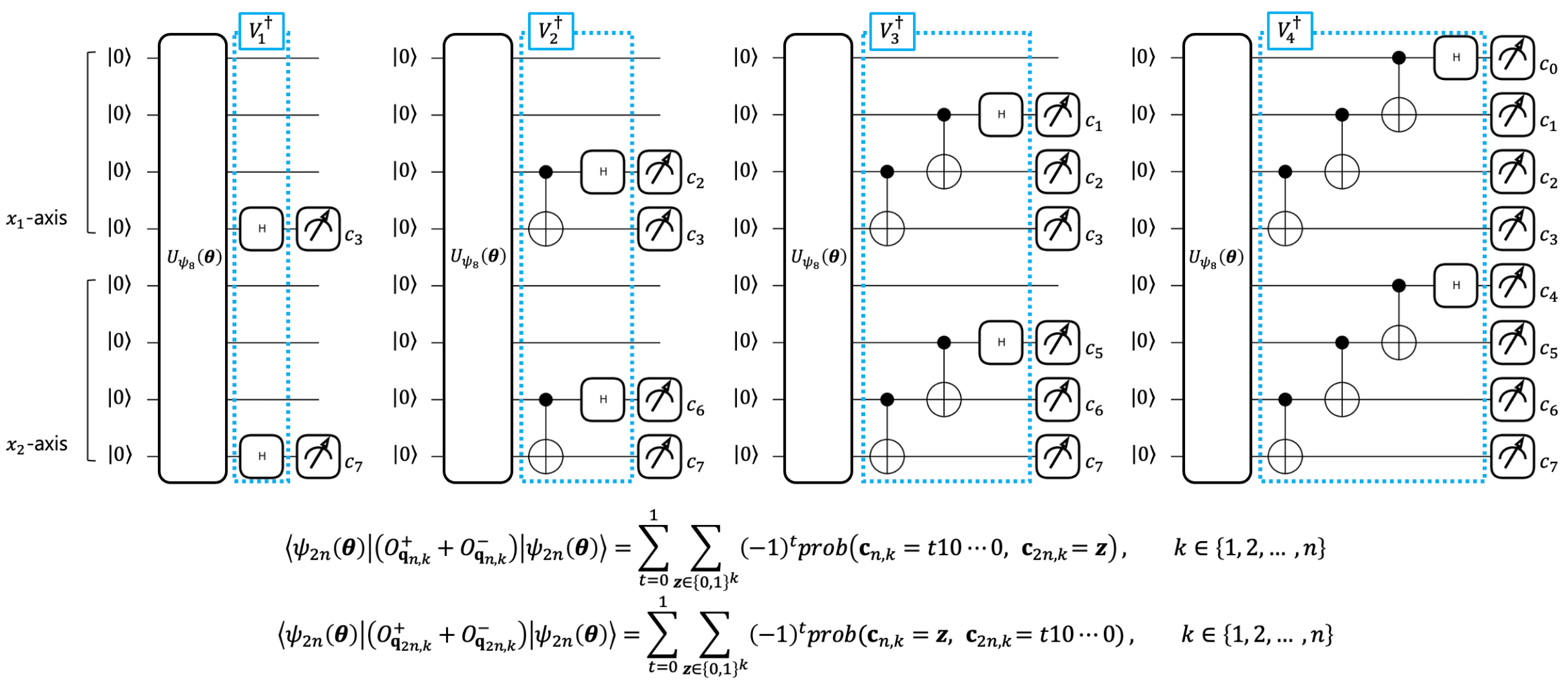}
\caption{\label{fig_denominator}
\textbf{Quantum circuits for the case of (\textit{d} = 2, \textit{n} = 4) (II):} The circuits used to evaluate the denominator $\bra{\psi_{2n}(\pmb{\theta})}A_{DD}\ket{\psi_{2n}(\pmb{\theta})}$ of the cost function $E(\pmb{\theta})$. For each $k \in \{1, 2, \dots, n\}$, $\bra{\psi_{2n}(\pmb{\theta})}(O^{+}_{\mathbf{q}_{n,k}}+O^{-}_{\mathbf{q}_{n,k}}) \ket{\psi_{2n}(\pmb{\theta})}$ and $\bra{\psi_{2n}(\pmb{\theta})}(O^{+}_{\mathbf{q}_{2n,k}}+O^{-}_{\mathbf{q}_{2n,k}}) \ket{\psi_{2n}(\pmb{\theta})}$ can be evaluated with a single $2n$-qubit circuit by applying the unitary $V_{k}^{\dagger}$, so the number of circuits required to compute the denominator is independent of the dimension of the Poisson equation ($d$). $c_{j}$ in subfigures indicates the outcome obtained by measuring the $j$-th qubit ($q_{j}$), and $\mathbf{c}_{r,s}=c_{r-s} \cdots c_{r-1}$ for $1 \le s \le r$.
}
\end{figure*}

The value of $\bra{\psi_{dn}(\pmb{\theta})}A\ket{\psi_{dn}(\pmb{\theta})}$, which serves as the denominator of the cost function and its partial derivatives, can be estimated by decomposing the Poisson matrix $A$ into operators that can be efficiently implemented with quantum circuits. We first review the decomposition method used by Liu \textit{et al.}~\cite{Liu21} for the matrix $A_{D}$ of the one-dimensional Poisson equation with Dirichlet boundaries, which is represented by
\begin{eqnarray}
\label{decomp_A}
A_{D} &=& 2I^{(2)}_{q_{0}} \otimes \cdots \otimes I^{(2)}_{q_{n-1}} \nonumber \\
&&- \sum_{k=1}^{n} I^{(2)}_{q_{0}} \otimes \cdots \otimes I^{(2)}_{q_{n-1-k}} \otimes (O^{+}_{\mathbf{q}_{n,k}}+O^{-}_{\mathbf{q}_{n,k}}),
\label{eq:13}
\end{eqnarray}
where $q_{j}$ represents the $j$-th qubit ($0 \le j \le n-1$), $\mathbf{q}_{r,s}=q_{r-s} \cdots q_{r-1}$ ($1 \le s \le r$), $O^{+}_{\mathbf{q}_{n,k}} \equiv \ket{01 \cdots 1}_{\mathbf{q}_{n,k}}\bra{10 \cdots 0}$, and $O^{-}_{\mathbf{q}_{n,k}} \equiv \ket{10 \cdots 0}_{\mathbf{q}_{n,k}}\bra{01 \cdots 1}$. Since $O^{+}_{\mathbf{q}_{n,k}}$ and $O^{-}_{\mathbf{q}_{n,k}}$ are not unitary operators, Liu \textit{et al.} employed an ancilla qubit to estimate the values of $\bra{\psi_n(\pmb{\theta})}O^{+}_{\mathbf{q}_{n,k}}\ket{\psi_n(\pmb{\theta})}$ and $\bra{\psi_n(\pmb{\theta})}O^{-}_{\mathbf{q}_{n,k}}\ket{\psi_n(\pmb{\theta})}$ (note that here $I^{(2)}_{q_{0}} \otimes \cdots \otimes I^{(2)}_{q_{n-1-k}}$ is omitted, and $\ket{\psi_n(\pmb{\theta})}$ is the normalized solution of the one-dimensional Poisson equation). 

The evaluation of $\bra{\psi_n(\pmb{\theta})}A_{D}\ket{\psi_n(\pmb{\theta})}$, however, can be done more efficiently with no ancilla qubits as $\bra{\psi_n(\pmb{\theta})}O^{+}_{\mathbf{q}_{n,k}}\ket{\psi_n(\pmb{\theta})}$ and $\bra{\psi_n(\pmb{\theta})}O^{-}_{\mathbf{q}_{n,k}}\ket{\psi_n(\pmb{\theta})}$ do not necessarily have to be obtained separately. Let us define $\ket{\phi_{k}^{t}}_{\mathbf{q}_{n,k}}$ as follows,
\begin{equation}
\ket{\phi_{k}^{t}}_{\mathbf{q}_{n,k}} = \frac{1}{\sqrt{2}}\left(\ket{0 1 \cdots 1}_{\mathbf{q}_{n,k}}
+(-1)^{t} \ket{1 0 \cdots 0}_{\mathbf{q}_{n,k}}\right)
\label{eq:14}
\end{equation}
for $t \in \{0, 1\}$.
Then we have
\begin{eqnarray}
&&\bra{\psi(\pmb{\theta})}(O^{+}_{\mathbf{q}_{n,k}}+O^{-}_{\mathbf{q}_{n,k}}) \ket{\psi(\pmb{\theta})} \nonumber \\
&&= \bra{\psi(\pmb{\theta})}
\left(\ket{\phi_{k}^{0}}_{\mathbf{q}_{n,k}}\bra{\phi_{k}^{0}}-\ket{\phi_{k}^{1}}_{\mathbf{q}_{n,k}}\bra{\phi_{k}^{1}}\right)
\ket{\psi(\pmb{\theta})},
\label{eq:15}
\end{eqnarray}
where $I^{(2)}_{q_{0}} \otimes \cdots \otimes I^{(2)}_{q_{n-1-k}}$ is omitted. Let us also define $V_{\mathbf{q}_{n,k}}$ as
\begin{equation}
V_{\mathbf{q}_{n,k}} \equiv \left(\prod_{i=1}^{k-1} CNOT_{q_{n-i-1},q_{n-i}}\right) H_{q_{n-k}},
\label{eq:16}
\end{equation}
where $H$ represents the Hadamard operator, and $CNOT_{q_{\alpha}, q_{\beta}}$ is the controlled-$X$ (CNOT) operator with a control qubit $q_{\alpha}$ and a target qubit $q_{\beta}$.
Then we have 
\begin{equation}
\ket{\phi_{k}^{t}}_{\mathbf{q}_{n,k}} = V_{\mathbf{q}_{n,k}}\ket{t 1 0 \cdots 0}_{\mathbf{q}_{n,k}}, 
\label{eq:17}
\end{equation}
and thus we can construct the measurement circuit to directly evaluate $\bra{\psi_n(\pmb{\theta})}(O^{+}_{\mathbf{q}_{n,k}}+O^{-}_{\mathbf{q}_{n,k}}) \ket{\psi_n(\pmb{\theta})}$. Note that here we need a total of $n$ quantum circuits to get the value of $\bra{\psi_n(\pmb{\theta})}A_{D}\ket{\psi_n(\pmb{\theta})}$ as it can be computed with $n$ terms except for the one consisting of identity operators, as shown in Equation (\ref{eq:13}).

The one-dimensional Poisson matrix with Neumann boundaries, $A_{N}$, satisfies
\begin{equation}
A_{N} = A_{D}-e_{1}e_{1}^{\dagger}-e_{m}e_{m}^{\dagger},
\label{eq:18}
\end{equation}
where $e_{1}$ and $e_{m}$ are $\ket{0 \cdots 0}_{\mathbf{q}_{n,n}}$ and $\ket{1 \cdots 1}_{\mathbf{q}_{n,n}}$, respectively. Since we can estimate the value of $\bra{\psi_n(\pmb{\theta})}(e_{1}e_{1}^{\dagger}+e_{m}e_{m}^{\dagger})\ket{\psi_n(\pmb{\theta})}$ by measuring $\ket{\psi_n(\pmb{\theta})}$ in the computational basis, the evaluation of $\bra{\psi_n(\pmb{\theta})}A_{N}\ket{\psi_n(\pmb{\theta})}$ requires one more quantum circuit than the case of Dirichlet boundaries. With a periodic boundary condition, the system matrix $A_{p}$ for the one-dimensional Poisson equation is written as 
\begin{equation}
A_{p}=A_{D}-e_{1}e_{m}^{\dagger}-e_{m}e_{1}^{\dagger},
\end{equation}
and the value of $\bra{\psi_n(\pmb{\theta})} (e_{1}e_{m}^{\dagger}+e_{m}e_{1}^{\dagger}) \ket{\psi_n(\pmb{\theta})}$ here can be obtained with the circuit that is utilized to evaluate $\bra{\psi_n(\pmb{\theta})}O^{+}_{\mathbf{q}_{n,n}}+O^{-}_{\mathbf{q}_{n,n}} \ket{\psi_n(\pmb{\theta})}$. Hence, $\bra{\psi_n(\pmb{\theta})}A_{p}\ket{\psi_n(\pmb{\theta})}$ can be calculated with $n$ quantum circuits. 

We now examine the Poisson matrix of a $d$-dimensional domain. If the domain has only Dirichlet boundaries, all the matrices $\{A_i\}$ in Equation (\ref{mat_dPoisson}) become $A_D$. Since $A$ has a total of $d\times n=dn$ terms except the one consisting of identity operators, as shown in Equation (\ref{mat_dPoisson}), $\bra{\psi_{dn}(\pmb{\theta})}A\ket{\psi_{dn}(\pmb{\theta})}$ can be evaluated with $dn$ quantum circuits if each term is described with one circuit. But, we can still estimate $\bra{\psi_{dn}(\pmb{\theta})}A\ket{\psi_{dn}(\pmb{\theta})}$ with $n$ quantum circuits since one-dimensional matrices $\{A_{i}\}$ for different axes are applied to different qubit groups in one circuit. For example, if we consider a two-dimensional Poisson matrix $A_{DD} = I^{(m)} \otimes A_{D} + A_{D} \otimes I^{(m)}$ and define $V_{k}$ as
\begin{equation}
V_{k} \equiv V_{\mathbf{q}_{n,k}} \otimes V_{\mathbf{q}_{2n,k}} (0 \le k \le n-1), 
\end{equation}
the values of $\bra{\psi_{2n}(\pmb{\theta})}(O^{+}_{\mathbf{q}_{n,k}}+O^{-}_{\mathbf{q}_{n,k}}) \ket{\psi_{2n}(\pmb{\theta})}$ and $\bra{\psi_{2n}(\pmb{\theta})}(O^{+}_{\mathbf{q}_{2n,k}}+O^{-}_{\mathbf{q}_{2n,k}}) \ket{\psi_{2n}(\pmb{\theta})}$ can be obtained with a single circuit, by applying the unitary operator $V_{k}^{\dagger}$ to the ansatz state $\ket{\psi_{2n}(\pmb{\theta})}$ as shown in Figure~\ref{fig_denominator}. Therefore, the number of quantum circuits required to estimate the value of $\bra{\psi_{2n}(\pmb{\theta})}A_{DD}\ket{\psi_{2n}(\pmb{\theta})}$ only depends on the number of decomposition terms for $A_{D}$.

The expectation value of a Poisson matrix can be obtained similarly when the domain has mixed boundary conditions. Let us consider a two-dimensional domain that has Neumann and Dirichlet boundaries along the $x_{1}$- and the $x_{2}$-axis, respectively. The corresponding Poisson matrix $A_{ND} = A_{N} \otimes I^{(m)} + I^{(m)} \otimes A_{D}$, and $A_{ND}$ can be written as 
\begin{equation}
A_{ND} = I^{(m)} \otimes A_{D} + A_{D} \otimes I^{(m)} -(e_{1}e_{1}^{\dagger}+e_{m}e_{m}^{\dagger})\otimes I^{(m)}.
\end{equation}
So, compared to the case of $A_{DD}$, we need one more circuit to estimate the value of $\bra{\psi_{2n}(\pmb{\theta})}(e_{1}e_{1}^{\dagger}+e_{m}e_{m}^{\dagger})\otimes I^{(m)}\ket{\psi_{2n}(\pmb{\theta})}$, which can be done by measuring qubits associated to the $x_{1}$-axis in the computational basis.

\section{Results and Discussion}
\label{Sec: results and discussion}

\subsection{Experiments with two-dimensional Poisson equations}
\label{sec:toy problems}

In VQAs, classical and quantum computing are utilized in a hybrid mode, so the cost function is evaluated with the ansatz state in a quantum computer as discussed in the section \ref{Sec: methods}, and the set of ansatz parameters are classically updated in an iterative manner. Here we mimic a gate-based quantum compute with the PENNYLANE software development kit~\cite{Pennylane}, which is a powerful and versatile quantum computing simulator, and update ansatz parameters with the well-known L-BFGS-B algorithm~\cite{Liu89} that can handle parameter spaces with memory-efficiency. To examine the functionality of our algorithm, we first employ a two-dimensional domain that is discretized with $m\times m$ grids. Taking two cases of boundary conditions into account, we describe the computational domain with a system matrix $A_{DD}$ and $A_{ND}$. For the RHS vector, we consider $\mathbf{b}=\ket{\gamma} \otimes \ket{\gamma}$ that represents a step function, where $\ket{\gamma}=[1, 1, \dots, 1, -1, -1, \dots, -1]/\sqrt{m}$. We note that $m$ = $2^n$, and $d$ = 2 in this case.

The vector $\mathbf{b}$ can be obtained with single-qubit unitaries in a 2$n$-qubit quantum circuit as follows,
\begin{equation}
\mathbf{b}=\left(Z_{q_{0}} \otimes Z_{q_{n}}\right)H^{\otimes 2n}\ket{0}^{\otimes 2n},
\end{equation}
where $Z$ represents the Pauli-$Z$ gate. To evaluate the similarity between the VQA-driven solution $\ket{\psi}$ and the normalized exact solution $\ket{\mathbf{u}}$ of the linear system problem, we compute their fidelity $F=|\inn{\psi}{\mathbf{u}}|$. Because a $L^2$ norm of the solution vector contains important information in many problems, it is crucial to examine if the VQA can provide a sufficiently accurate norm. The accuracy of the norm obtained with our VQA is assessed with its relative error against the precise value that can be calculated as
\begin{equation}
\left|\frac{||A^{-1}\mathbf{b}||-|r(\pmb{\theta}_{opt})|||\mathbf{b}||}{||A^{-1}\mathbf{b}||}\right|,
\end{equation}
where the Poisson matrix $A$ becomes either $A_{DD}$ or $A_{ND}$ in this case-study.

\begin{figure}[t]
\includegraphics[width=\columnwidth]{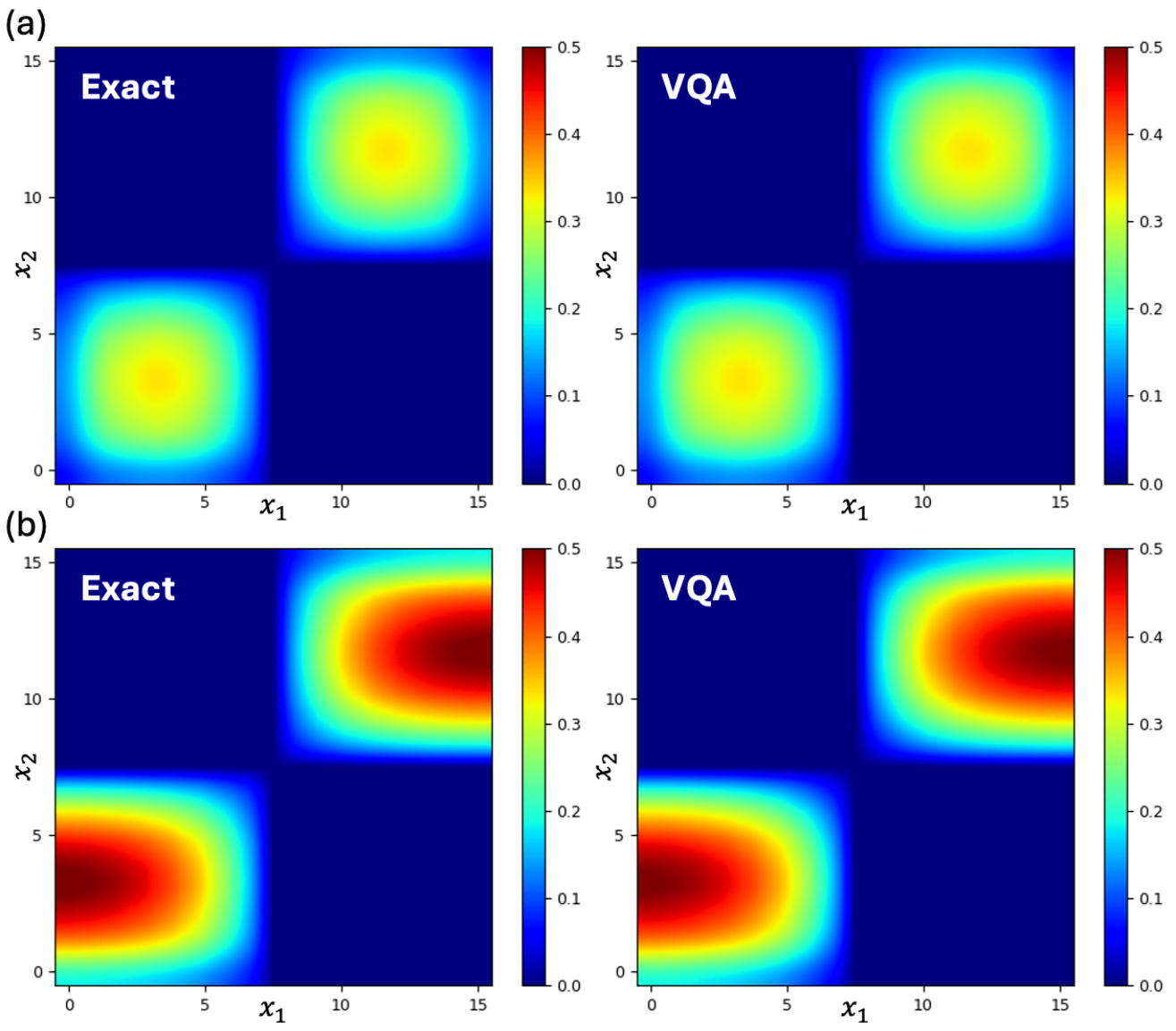}
\caption{\label{fig_toy}
\textbf{Simulation results of the two-dimensional problem with a step function RHS:} (a) The exact solution (left) and the VQA-driven solution when the domain has only Dirichlet boundaries (the system matrix = $A_{DD}$), (b) The exact solution (left) and the VQA-driven solution when the domain has Neumann and Dirichlet boundaries (the system matrix = $A_{ND}$). The domain is discretized with 16$\times$16 grids so $n=4$ and $d=2$, and simulations are conducted without consideration of noise or sampling errors. For both cases, VQA-driven results show high fidelities ($>$ 0.9999) with low norm errors ($\simeq$ 0.0007).
}
\end{figure}

\begin{figure*}[t]
\includegraphics[width=\textwidth]{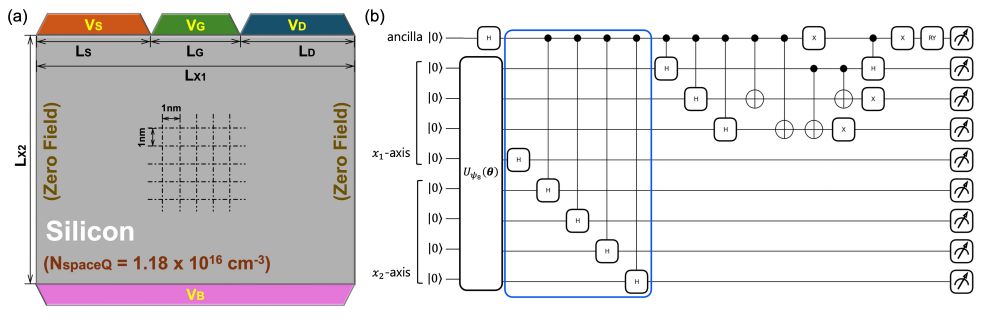}
\caption{\label{fig_device_problem}
\textbf{The problem of finding potential distributions in a simple semiconductor system:} (a) Illustration of the two-dimensional simulation domain. The system is based on silicon material that has a constant space charge density ($N_{spaceQ}$) = 1.18$\times10^{16}\rm{cm}^{-3}$. Being discretized with 1nm $\times$ 1nm rectangular grids, the domain has three electrodes on the top and one electrode on the bottom. Lengths of the top leads ($L_{S}$, $L_{G}$, $L_{D}$) satisfy the condition of $2L_{S}=3L_{G}=2L_{D}$. Numerical simulations are conducted with $V_{G}$ = (0.1$V$, 0.3$V$, 0.5$V$), while all the other leads are assumed to be grounded ($V_S$ = $V_D$ = $V_B$ = 0). Electric fields at boundaries along the $x_1$-axis are assumed to be zero. (b) A quantum circuit, which can evaluate the numerator of the cost function based on the Equation (\ref{eq:27}), is shown when $n=4$ where the quantum gate applied at the final stage is $RY(-2\eta)$. We note that changing $n$ affects only the part enclosed with a blue solid line, where the number of associated gates is linearly proportional to $n$.
}
\end{figure*}

In Figure~\ref{fig_toy}, we show the solutions that are obtained with VQA in an ideal condition where circuit operations are not affected by noises and sample errors. Numerical simulations are conducted for a two-dimensional domain described with 16$\times$16 grids ($n=4$). The ansatz circuit shown in Figure \ref{fig_ansatz}(a) is constructed by repeating the part enclosed with a red dotted line three times (let us say this number of repetition = $p$), and the initial set of ansatz parameters ($\pmb{\theta}_{opt}$) are determined randomly. In both cases of boundary conditions, the VQA-driven solutions are quite accurate so their fidelities become larger than 0.9999 and corresponding relative errors of $L^2$ norms are about 0.0007. Results here clearly highlight the reliability of our VQA for securing solutions of multi-dimensional Poisson equations with mixed boundary conditions in an ideal condition, establishing a solid foundation for moving forward to a bit more complicated and realistic problems.

\subsection{Calculation of potential distributions in a semiconductor system having multiple electrodes}
\label{sec:target problem}

One of computational tasks that requires the Poisson equation is to predict the spatial distribution of electric fields under a given profile of charge densities. In particular, engineering of semiconductor devices involves rigorous evaluation of electric fields since they not only determine current-voltage characteristics of traditional devices like field-effect transistors \cite{Liou1998, Amiri2019} and junction-based diodes \cite{Khan1987, Milos2009}, but significantly affect confinement of carriers in devices of a nanometer(nm)-scale like nanowire transistors \cite{Ryu2016_2, Neophytos2011} and quantum dot structures designed for optoelectronic devices \cite{Usman2011, Ahmed2015} or quantum processing units \cite{Ryu2022, Kang2021_2}. Upon what have been discussed so far, here we apply our VQA to calculations of bias-dependent potential distributions in a simple semiconductor system that has multiple contacts. Figure~\ref{fig_device_problem}(a) illustrates the problem we target to solve, where we define a two-dimensional domain of silicon with a space charge density of 1.18 $\times$ 10$^{16}$cm$^{-3}$. Being discretized with 1nm $\times$ 1nm rectangular grids, the system has four electrodes such that three of them are on top with lengths of ($L_S$, $L_G$, $L_D$), and the remaining one is placed on bottom with a length that is same as the dimension of the domain along the $x_{1}$-axis. The lengths of three top electrodes satisfy the condition of 2$L_S$ = 3$L_G$ = 2$L_D$, and the bias imposed on the middle electrode on top ($V_G$) is varied from 0.1V to 0.5V with a step of 0.2V, while other three electrodes are grounded ($i.e.$ a zero voltage is applied). The potential distribution is assumed to be flat ($i.e.$ a zero electric field) at boundaries along the $x_1$-axis. 

\begin{figure*}[t]
\includegraphics[width=\textwidth]{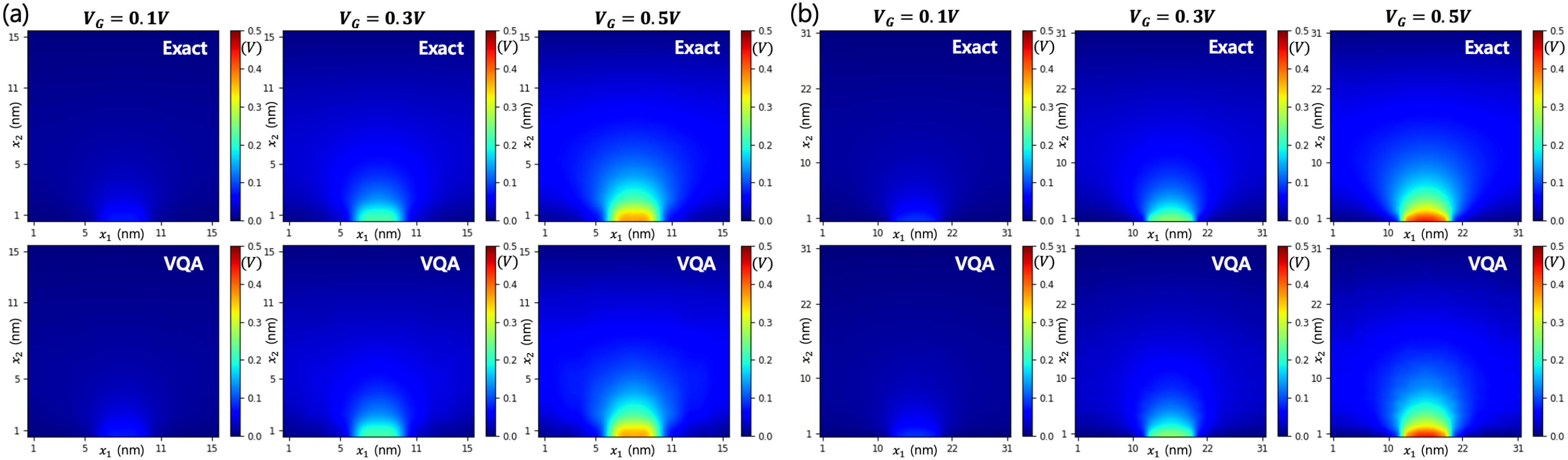}
\caption{\label{fig_device_results}
\textbf{Bias-dependent potential distributions:} (a) Exact solutions (top) and VQA-driven solutions (bottom) that are obtained in an ideal condition for a 16nm$\times$16nm domain ($n=4$) with 5 repeated blocks of the ansatz circuit ($p=5$) given in Figure \ref{fig_ansatz}(a). Results show fidelities exceeding 0.999 with $L^2$ norm errors $<$ 0.009 regardless of $V_G$. (b) Results obtained in an ideal condition for a 32nm$\times$32nm domain ($n=5$) with 6 repeated blocks of the ansatz circuit ($p=6$). VQA-driven solutions still have high fidelities $>$ 0.997 with norm errors lower than 0.05.
}
\end{figure*}

Clearly, the two-dimensional Poisson matrix in this case becomes $A_{ND}$ since we assume Neumann and Dirichlet boundaries along the $x_1$- and the $x_2$-axis, respectively. The RHS $\mathbf{b}$ is given by
\begin{equation}
\mathbf{b} = \alpha_{0} \mathbf{b}_{0} + \alpha_{1} \mathbf{b}_{1},
\label{eq:24}
\end{equation}
where $\alpha_{0} \in \{$0.1V, 0.3V, 0.5V$\}$, $\alpha_{1} \simeq 1.810 \times 10^{-5}$V, and
\begin{eqnarray}
\mathbf{b}_{0}& = &\sqrt{2}^{n-3}\left(\ket{011}+\ket{100}\right) \otimes \ket{+}^{\otimes n-3} \otimes \ket{0}^{\otimes n}, \nonumber \\
 \mathbf{b}_{1}& = &\sqrt{2}^{2n}\ket{+}^{\otimes 2n}.
\end{eqnarray}
We note that $\alpha_1$ in Equation (\ref{eq:24}) is calculated as $\epsilon_{si}^{-1}$ $\times$ ($\Delta x$)$^2$ $\times$ $Q$ $\times$ $N_s$, where $\epsilon_{si}$ is the permittivity of bulk silicon ($\simeq$ 1.044 $\times$ 10$^{-10}$F$/$m), $\Delta x$ is the size of discretization (= 10$^{-9}$m), Q is the charge of a single proton ($\simeq$ 1.602 $\times$ 10$^{-19}$C), and $N_s$ indicates the space charge density (= 1.180 $\times$ 10$^{22}$m$^{-3}$). In this case, it is challenging to find an efficient way to represent the unitary $U_{\mathbf{b}}$ with a 2$n$-qubit quantum circuit. While the method developed by Mottonen {\em et al.}~\cite{mottonen04} can be employed, the result may be inefficient in terms of the computing cost as we need $\sim$$O(2^{2n})$ CNOT and single-qubit gates to represent $U_{\mathbf{b}}$. To compute the numerator of $E(\pmb{\theta})$, therefore, we introduce an ancilla qubit and define the state $\ket{\mathbf{b}'}$ as follows,
\begin{equation}
\label{eq:26}
\ket{\mathbf{b}'}=\cos\eta\ket{0}\otimes\ket{\mathbf{b}_{0}} + \sin\eta\ket{1}\otimes\ket{\mathbf{b}_{1}},
\end{equation}
where $\ket{\mathbf{b}_{0}}$ and $\ket{\mathbf{b}_{1}}$ are the normalized version of $\mathbf{b}_{0}$ and $\mathbf{b}_{1}$, respectively, and $\eta = \tan^{-1}(\sqrt{2}^{n+2}\alpha_{1}/\alpha_{0})$. Now, the numerator of the cost function can be easily calculated with
\begin{equation}
\label{eq:27}
\inn{\mathbf{b}}{\psi_{2n}(\pmb{\theta})}^{2}=2(\beta_{0}^{2}+\beta_{1}^{2})\left(\bra{\mathbf{b}'}\left(\ket{+}\otimes\ket{\psi_{2n}(\pmb{\theta})}\right)\right)^{2},
\end{equation}
where $\beta_{i}=\alpha_{i}||\mathbf{b}_{i}||/||\mathbf{b}||$ ($i=0, 1$). It should be noted here that the $(\bra{\mathbf{b}'}\left(\ket{+}\otimes\ket{\psi_{2n}(\pmb{\theta})}\right))^{2}$ term can be evaluated with a quantum circuit that has the depth of $O(n)$, and the corresponding circuit when $n = 4$ is depicted in Figure~\ref{fig_device_problem}(b). 

\begin{figure}[t]
\includegraphics[width=\columnwidth]{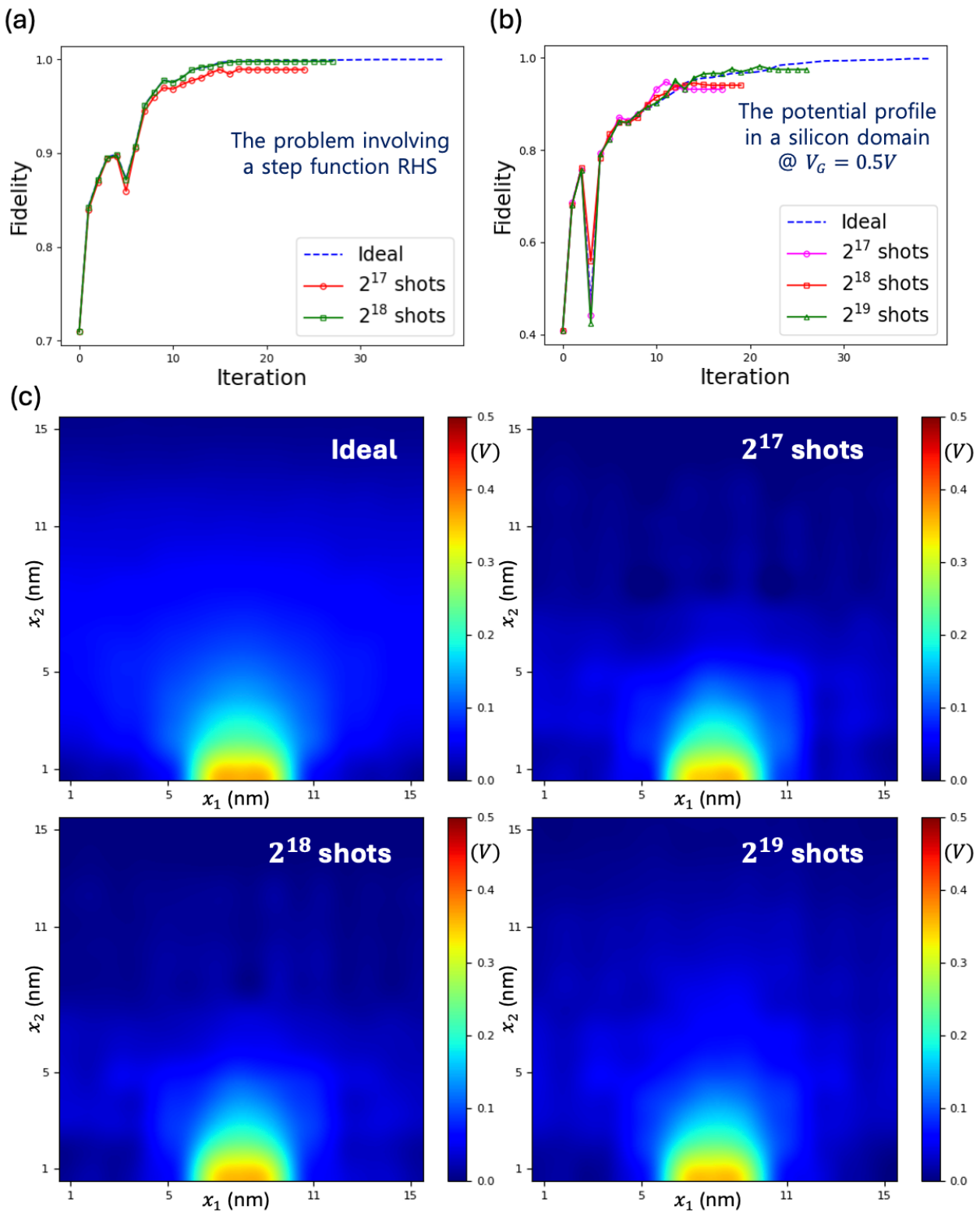}
\caption{\label{fig_samplingerror}
\textbf{Sampling errors:} (a) The convergence pattern of the VQA-driven solution is examined for the problem involving a step function RHS ($V_G=0.5V$ \& mixed boundaries). With $2^{17}$ shots, the fidelity and the norm error of the solution are 0.989 and 0.1, respectively. With $2^{18}$ shots, they increase to $0.998$ and $0.001$, respectively, being comparable to the ideal result. (b) The convergence pattern of a bias-dependent potential distribution ($V_G=0.5V$) is shown. The pattern here is generally worse than the case of Figure \ref{fig_samplingerror}(a) since the problem described in the section \ref{sec:target problem} has more complex RHS vectors, so, when $2^{19}$ shots are applied, the solution has a fidelity of 0.975 and its $L^2$ norm error $\simeq$ 0.18). (c) The converged potential profile at $V_G = 0.5V$ is shown as a function of the sampling number. While the fidelity mark 0.975 with $2^{19}$ shots, the potential distribution near the top middle electrode (the region of large electric field) is fairly close to the ideal result.}
\end{figure}

Figures~\ref{fig_device_results}(a) and~\ref{fig_device_results}(b) show the results of numerical simulations that are obtained with our VQA in an ideal condition for a 16nm $\times$ 16nm ($n = 4$) and a 32nm $\times$ 32nm structure ($n = 5$), respectively. Here the VQA-driven solutions become fairly good enough to claim their accuracy with ansatz circuits of $p = 5$ and 6 in the cases of $n = 4$ and 5, respectively. When $n = 4$, bias-dependent potential profiles have fidelities $>$ $0.999$ with relative norm errors $<$ $0.01$. For the case of a larger domain where $n = 5$, solutions have fidelities $>$ $0.997$, showing relative norm errors $<$ $0.05$. Here, the RHS vectors are more complicated than what we considered in the section \ref{sec:toy problems}, and the ansatz circuit requires more repeated blocks (enclosed with a red dotted line in Figure \ref{fig_ansatz}(a)) to secure accurate solutions. Though it is true   that results in Figure~\ref{fig_device_results} indicate the utility of VQA for tackling multi-dimensional physical problems with the Poisson equation, we can still expect that, in general, more complex problems may require larger spaces of parameters that are subjected to optimization. A more important message we can drive here is that the real practicality of VQA for linear system problems including Poisson equations would significantly depend on the shape of RHS vectors. We therefore need to carefully examine first whether given physical problems have RHS vectors that can be implemented with quantum circuits in a cost-efficient manner.

\subsection{Effects of sampling errors on VQA performance}
\label{sampling errors}
So far, we have conducted numerical simulations under an ideal condition where effects of noises or sampling errors are ignored. In NISQ devices, however, circuit operations cannot be free from noises, and the cost function must be evaluated through multiple measurements since the information of each qubit is always collapsed to either $\ket{0}$ or $\ket{1}$ upon a single measurement. As the first step to examine the performance of our VQA in more realistic conditions, here we conduct numerical simulations with sampling effects for the two-dimensional problems of mixed boundaries that are discussed the sections \ref{sec:toy problems} and \ref{sec:target problem}. For the problem involving a silicon structure (Figure \ref{fig_device_problem}(a)), we only test the case where $V_g$ = 0.5V is imposed on the middle electrode on top of the domain. For both problems, we consider the case of $n=4$, so 8-qubit circuits are employed to describe the Poisson domains. Numerical simulations are conducted with 2$^{17}$- 2$^{19}$ shots, where the maximal number of shots in our case is similar to what was considered (= $10^6$) by Sato {\em et al.} for one-dimensional problems~\cite{Sato21}.

In Figure~\ref{fig_samplingerror}(a), we show the results of the two-dimensional Poisson problem that has a step-function-like RHS. Here, the VQA-driven solution becomes closer to the ideal result as we conduct more shots, so the fidelity with $2^{17}$ and $2^{18}$ shots reach 0.989 and 0.998 with a $L^2$ norm error of $0.1$ and $0.001$, respectively, where the fidelity and norm error are 0.9999 and 0.0007 in the ideal case. The problem of securing potential profiles in the silicon structure that has more complicated RHS vectors, however, generally needs more samples to make the VQA solutions close to the ideal results as shown in Figure~\ref{fig_samplingerror}(b). Ideally, the fidelity is about $0.999$ and the norm error is $\sim$$0.01$. With $2^{17}$, $2^{18}$, and $2^{19}$ shots, however, the fidelity and corresponding $L^2$ norm error become (0.932, 0.23), (0.941, 0.22), and (0.975, 0.18), respectively. As we see from Figure~\ref{fig_samplingerror}(c) that shows the simulated two-dimensional potential profile as a function of the sampling number, the result with a fidelity of 97.5\% looks similar to the desired profile in the vicinity of the electrode where 0.5V is imposed, but the difference gets remarkable as we move farther towards other boundaries where the potential varies in small magnitudes. We remind you this phenomenon has nothing to do with noises but solely depends on the sampling error that would happen even in fault-tolerant quantum computers, indicating strong needs of efficient strategies for the reconstruction of states with a reasonable number of samples.  

\subsection{Circuit complexity and noise dependency}
\label{comparison}

\begin{figure}[t]
\includegraphics[width=\columnwidth]{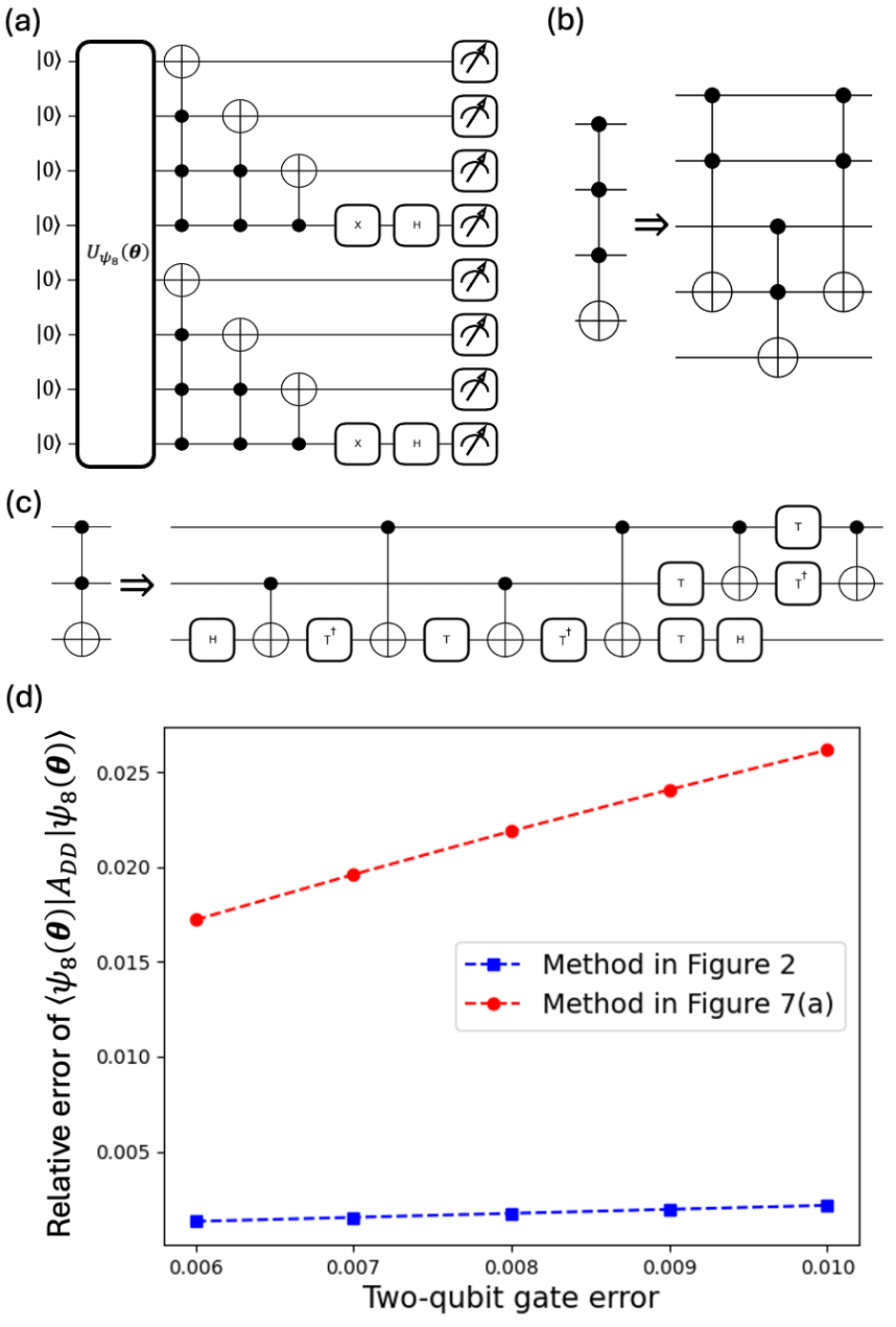}
\caption{\label{fig_comparison}
\textbf{The strength of our algorithm:}
(a) The quantum circuit for estimating the expectation value $\bra{\psi_{dn}(\pmb{\theta})}A_{DD}\ket{\psi_{dn}(\pmb{\theta})}$ with multi-controlled $X$ gates, based on the decomposition scheme proposed by Sato {\em et al.}~\cite{Sato21}, is shown for the case $d=2$ and $n=4$. (b) A single CCCX gate can be decomposed into a combination of Toffoli gates with an ancilla qubit, and (c) a single Toffoli gate can be decomposed into one- and two-qubit gates. (d) When we calculate the expectation value using the circuit in Figure \ref{fig_comparison}(a) that is decomposed with the schemes in Figures \ref{fig_comparison}(b) and \ref{fig_comparison}(c), the result is more sensitive to depolarizing errors than the one calculated with our method shown in Figure~\ref{fig_denominator}. Note that here we show the mean values of $30$ trials as a function of the two-qubit gate error, where $\pmb{\theta}$ is randomly determined in each trial.
}
\end{figure}

To solve $d$-dimensional Poisson equations that employ $m=2^n$ grids to discretize the domain along a single axis, one may consider the following cost function
\begin{equation}
\tilde{E}(\pmb{\theta}) \equiv \bra{\psi_{dn}(\pmb{\theta})}A^{2}\ket{\psi_{dn}(\pmb{\theta})}-|\bra{\mathbf{b}}A\ket{\psi_{dn}(\pmb{\theta})}|^{2},
\end{equation}
which is derived from the problem of finding the ground state energy of $H \equiv A^{\dagger}A-A^{\dagger}\ket{\mathbf{b}}\bra{\mathbf{b}}A$~\cite{Bravo23}, and was used by Liu $et$ $al$. for one-dimensional problems~\cite{Liu21}. Evaluation of $\tilde{E}(\pmb{\theta})$ however requires more quantum circuits than what are needed to evaluate $E(\pmb{\theta})$ given in Equation (\ref{eq:08}), particularly due to the term of $\bra{\psi_{dn}(\pmb{\theta})}A^{2}\ket{\psi_{dn}(\pmb{\theta})}$. For a two-dimensional problem that has only Dirichlet boundaries, the Poisson system matrix $A$ becomes $A_{DD}=I^{(m)} \otimes A_{D} + A_{D} \otimes I^{(m)}$, so $A_{DD}^{2}=I^{(m)} \otimes A_{D}^{2} + A_{D}^{2} \otimes I^{(m)} + A_{D} \otimes A_{D}$. Now, $A_{D}^2$ can be written as
\begin{eqnarray}
\label{decomp_A2}
A_{D}^{2} &=& 6I^{(2)}_{q_{0}} \otimes \cdots \otimes I^{(2)}_{q_{n-1}} -(e_{1}e_{1}^{\dagger}+e_{m}e_{m}^{\dagger})\nonumber \\
&& -4 \sum_{k=1}^{n-1} I^{(2)}_{q_{0}} \otimes \cdots \otimes I^{(2)}_{q_{n-2-k}} \otimes (O^{+}_{\mathbf{q}_{n-1,k}}+O^{-}_{\mathbf{q}_{n-1,k}}) \otimes I^{(2)}_{q_{n-1}} \nonumber \\
&& -4 \sum_{k=1}^{n} I^{(2)}_{q_{0}} \otimes \cdots \otimes I^{(2)}_{q_{n-1-k}} \otimes (O^{+}_{\mathbf{q}_{n,k}}+O^{-}_{\mathbf{q}_{n,k}}),
\end{eqnarray}
where $O^{\pm}_{\mathbf{q}_{n,k}}$, $e_{1}$ and $e_{m}$ are defined in Equations (\ref{eq:13}) and (\ref{eq:18}). Since the expectation value of ($O^{+}_{\mathbf{q}_{n,k}}+O^{-}_{\mathbf{q}_{n,k}}$) and ($e_{1}e_{1}^{\dagger}+e_{m}e_{m}^{\dagger}$) can be evaluated with a single circuit as discussed in the section \ref{Quantum circuits}, the expectation value of $A_{D}^{2}$ can be evaluated with $2n$ quantum circuits. The expectation value of $A_{D} \otimes A_{D}$ can be computed with $n^2$ circuits since $A_{D}$ has $n$ terms except one term involving a string of identity operators as shown in Equation (\ref{eq:13}). Therefore, $\bra{\psi_{2n}(\pmb{\theta})} A_{DD}^{2} \ket{\psi_{2n}(\pmb{\theta})}$ in $\tilde{E}(\pmb{\theta})$ can be computed with a total of $n^{2}+2n$ quantum circuits, and the cost-efficiency of our method is obvious since we utilize $n$ circuits to compute $\bra{\psi_{2n}(\pmb{\theta})}A_{DD}\ket{\psi_{2n}(\pmb{\theta})}$ in $E(\pmb{\theta})$. 

The cost function employed by Sato {\em et al.} is basically identical to ours~\cite{Sato21}. They however decompose the Poisson system matrices with multi-controlled $X$ gates as illustrated in Figure~\ref{fig_comparison}(a), and thus $\bra{\psi_{2n}(\pmb{\theta})}A_{DD}\ket{\psi_{2n}(\pmb{\theta})}$ can be evaluated with a constant number of quantum circuits. Since we need $n$ quantum circuits for the same task, the decomposition scheme proposed by Sato {\em et al.} seems to be advantageous if we focus on the number of required quantum circuits. The utilization of multi-controlled $X$ gates in NISQ devices, however, can cause substantial errors driven by noises, since they need to be realized by circuit decompositions based on single- and two-qubit gates that can be directly supported by devices. Figures~\ref{fig_comparison}(b) and \ref{fig_comparison}(c) illustrate how a single four-qubit CCCX gate, which is used to get the expectation value of a two-dimensional Poisson matrix with $n=4$, is decomposed with single- and two-qubit gates, where it is assumed NISQ devices directly support CNOT operations for any qubit pairs. Even though we have assumed the feasibility of direct CNOT operations for any two qubits, a single CCCX gate still requires a circuit of depth = 36, and the depth of a circuit in Figure~\ref{fig_comparison}(a) becomes 51 excluding the ansatz and measurement operators, being much larger than those of our circuits in Figure \ref{fig_denominator}. Consequently, our method can be still advantageous in terms of noise robustness of circuit operations.

To examine the robustness of noisy operations in more detail, we decompose the circuit in Figure \ref{fig_comparison}(a) into one- \& two-qubit gates, and evaluate $\bra{\psi_{2n}(\pmb{\theta})}A_{DD}\ket{\psi_{2n}(\pmb{\theta})}$ with depolarizing channel errors. After conducting the same task using our circuits given in Figure \ref{fig_denominator}, we compute the relative error of the expectation value as follows
\begin{equation}
\left | \frac{\bra{\psi_{2n}(\pmb{\theta})}A_{DD}\ket{\psi_{2n}(\pmb{\theta})}_{ideal} - \bra{\psi_{2n}(\pmb{\theta})}A_{DD}\ket{\psi_{2n}(\pmb{\theta})}_{noisy}}{\bra{\psi_{2n}(\pmb{\theta})}A_{DD}\ket{\psi_{2n}(\pmb{\theta})}_{ideal}} \right |.
\end{equation}

In the 127-qubit IBM quantum computer, the median error rate of two-qubit logics is about 2.40$\times$$10^{-4}$, while that of single-qubit gates is $\simeq$ 8.11$\times$$10^{-3}$~\cite{IBM}. To get relative errors, therefore, we consider the error rate of two-qubit gates from 6.0$\times$$10^{-3}$ to 1.0$\times$$10^{-2}$, fixing the error rate of single-qubit operations to 2.0$\times$10$^{-4}$. Simulation results in Figure~\ref{fig_comparison}(d) clearly indicate that employing multi-controlled $X$ gates would not be a good choice, implying the strength of our method in NISQ devices.

\section{Conclusion}
\label{sec: conclusion}

We have designed a variational quantum algorithm (VQA) to discuss its practicality for securing solutions of the Poisson equations that describe multi-dimensional computational domains with mixed boundary conditions. Adopting a cost function based on the concept of the minimal potential energy, we have proposed the strategy for evaluation of the cost function that is more beneficial in terms of the circuit complexity and the noise robustness than what have been reported by previous studies for handling one-dimensional toy problems. To examine the practicality of our strategy, we have conducted numerical experiments against two-dimensional problems using the PENNYLANE simulator, including the one that computes the spatial distribution of potential energy in lightly doped silicon that are dependent on electrical biases imposed on multiple electrodes. Simulation results not only show the solutions are fairly accurate in an ideal condition, but also confirm that the expectation value of a Poisson system matrix can be computed with much less sensitivity to depolarizing noises than the case of previous studies. Though here we have shown the solid applicability of VQA to Poisson equations, we still want to note that the effective practicality of VQA for handling linear system problems can substantially depend on the right-hand-side (RHS) vectors that may vary if we change boundary conditions for the same domain or target problems involving different domains. Consequently, if one plans to solve linear system problems including Poisson equations with VQA, it should be crucial to examine in advance whether associated RHS vectors can be represented with quantum circuits in a cost-effective manner.


\section*{ACKNOWLEDGMENTS}
This research work has been carried out under the support from the National Research Foundation of Korea (NRF) grant (NRF-2022M3K2A1083890) that is funded by the Korea government (MSIT).


\bibliography{main}

\end{document}